\documentclass{article}
\usepackage{kr}

\usepackage{times}
\usepackage{soul}
\usepackage{url}
\usepackage[hidelinks]{hyperref}
\usepackage[utf8]{inputenc}
\usepackage[small]{caption}
\usepackage{graphicx}
\usepackage{amsmath}
\usepackage{amsthm}
\usepackage{booktabs}
\usepackage{algorithm}
\usepackage{algorithmic}
\usepackage{mathtools}
\usepackage{todonotes}

\newtheorem{theorem}{Theorem}
\newtheorem{lemma}{Lemma}
\newtheorem{corollary}{Corollary}[theorem]

\theoremstyle{definition}
\newtheorem{definition}{Definition}[section]
\usepackage{amsfonts}

\usepackage{tikz}

\title{Verification and Realizability in Finite-Horizon Multiagent Systems}

\author{%
Senthil Rajasekaran and
Moshe Y. Vardi
\affiliations
Rice University\\
\emails
\{sr79, vardi\}@rice.edu
}

\begin{document}

\maketitle
\begin{abstract}
    
The problems of \emph{verification} and \emph{realizability} are two central themes in the analysis of reactive systems. When multiagent systems are considered, these problems have natural analogues of existence (nonemptiness) of pure-strategy Nash equilibria and verification of pure-strategy Nash equilibria. Recently, this body of work has begun to include finite-horizon temporal goals. With finite-horizon temporal goals, there is a natural hierarchy of goal representation, ranging from deterministic finite automata (DFA), to nondeterministic finite automata (NFA), and to alternating finite automata (AFA), with a worst-case exponential gap between each successive representation. Previous works showed that the realizability problem with DFA goals was PSPACE-complete, while the realizability problem with temporal logic goals is in 2EXPTIME. In this work, we study both the realizability and the verification problems with respect to various goal representations. We first show that the realizability problem with NFA goals is EXPTIME-complete and with AFA goals is 2EXPTIME-complete, thus establishing strict complexity gaps between realizability with respect to DFA, NFA, and AFA goals. We then contrast these complexity gaps with the complexity of the verification problem, where we show that verification with respect to DFAs, NFA, and AFA goals is PSPACE-complete.
\end{abstract}

% \pagestyle{fancy}
% \fancyhead{}
% \maketitle 
\section{Introduction}
\emph{Verification} \cite{CES86} and \emph{Realizability} \cite{PR89a} are two major decision problems in the study of reactive systems. When the goals of these systems are specified through linear temporal logics, game theory has provided a powerful modeling framework for both problems through a two-agent game in which one agent takes on the role of a system that tries to realize a property and the other takes on the role of the environment that tries to falsify the property. The verification problem corresponds to checking whether an input strategy is winning for the system agent in the relevant game \cite{KV96a}, and the realizability (also called \emph{nonemptiness}) problem corresponds to determining whether a winning strategy for the system agent  exists \cite{PR89a}.

When the number of autonomous agents increases, the games become \emph{concurrent multiagent games}, suitable for analyzing concurrent multiagent systems \cite{SLmultiagentbook}.  In this setting, the notion of a winning strategy no longer corresponds to a meaningful solution concept, as there are no longer only two agents with a  purely adversarial relationship. In these types of systems, the concept of a \emph{pure-strategy Nash equilibria} (henceforth, Nash equilibria) has come to be a widely used solution concept \cite{BBMU15}. Informally, a Nash equilibria is a profile of strategies such that for each agent in the system deviating from the profile is never more profitable than not . In this sense, Nash equilibria represent a stable point that games naturally tend towards over repeated play \cite{Nash48,folktheorem}.

Concurrent multiagent games represent an extremely broad class of games.  \emph{Iterated Boolean Games} \cite{iBG} are a restriction of concurrent multiagent games that naturally mirror the games that model the two-agent realizability and verification problems \cite{KV96a,PR89a}. In an iterated boolean game each agent has a temporal goal and at each time step assigns a setting to a unique collection of boolean variables under its control. Thus, when all agents' assignments are considered we are given a complete valuation of the boolean variables at each time step. This infinite sequence of valuations is then used to determine which temporal goals are satisfied and which are not. Finding Nash equilibria in such games corresponds to a useful method of analysis of the systems that these games model; as such, there is a very large of body finding Nash equilibria when agents' goals are given by an infinite-horizon logic such as \emph{Linear Time Temporal Logic} (LTL) \cite{Wool11,GutierrezNPW20,MMPV14,GTW02,AGHKNPSW21}. 

Some systems, however, are naturally modeled by agents with finite-horizon goals, such as when notions like `completion' are considered. The concept of a finite-horizon temporal logic remains a relatively recent development in the study of temporal logics \cite{GV13}. While agents still create an infinite trace by setting their variables at every time step, satisfaction is considered over finite prefixes. The analogous problem of finding Nash equilibria in iterated boolean games in which each agent has been given a finite-horizon temporal goal has recently begun to receive attention \cite{RV21,GPW17}. 

Our modelling of this problem is done from the viewpoint of a system designer. Specifically, when given a system in which multiple agents have finite-horizon temporal logic goals, we query a subset $W$  of ``good'' agents to see if there is Nash equilibrium in which precisely the agents in $W$ are able to satisfy their goals. By the definition of the Nash equilibrium, this means that agents not within $W$, which  we consider as ``bad'' agents, are unable to unilaterally change their strategy and satisfy their own ``bad'' goal. In doing so we can naturally incorporate malicious agents with goals contrary to the designer's intent by specifying a set $W$ that not contain such agents. This study of teams of cooperating agents has clear parallels in earlier work in rational synthesis \cite{FKL10,KupfermanPV16}.

Here we consider \emph{Linear Time Temporal Logic on Finite Traces} (LTLf) \cite{GV13} as our standard finite-horizon temporal logic, but by using an automata-based approach we are able to prove more general results that are independent of a specific logic by considering the size of the automata that represents the specification.
Since finite-horizon temporal logics describe languages of finite words, they admit a variety of equivalent representations, ranging from deterministic finite automata (DFAs) to nondeterministic finite automata (NFAs) to alternating finite automata (DFAs) \cite{GV13,GV15,GV16}. While alternating finite automata are polynomial in the size with respect to their corresponding $LTL_f$ formula, nondeterminstic automata are exponential and deterministic finite automata are doubly exponential \cite{GV13}. By reasoning about different types of input automata, we are able to reason broadly about finite-horizon temporal goals with different goal succinctness 
%input AFAs being equivalent to inputs from the logic itself 
from a complexity-theoretic viewpoint.  It is then natural to consider how the succinctness of the representation influences the complexities of the realizability and verification problems. 

%It is here that we observe some very interesting and unexpected behavior. 
Our investigation sheds new light on the computational complexity of temporal Nash Equilibria. Note that, in prior work, the verification problem is usually proven to be easier than the realizability problem from a complexity-theoretic viewpoint. This corresponds to our intuition, since the verification problem checks a single input candidate strategy, while the realizability problem tries to find some solution strategy. Here, we observe the same phenomena of realizability being more difficult than verification. The succinctness of the representation does not, however, seem to effect  the complexity of the verification problem in this setting. No matter which representation we use, we get a PSPACE-complete complexity result. In contrast, if we consider the realizability problem then we get a strict hierarchy. For DFA goals, the problem is PSPACE-complete \cite{RV21} (the same as verification, an exception to the intuition that verification is easier than realizability). From NFAs goals, the problem is EXPTIME-complete. Finally, for AFA goals, the problem is 2EXPTIME-complete. This analysis extends the state of the art to include a complete set of results for both problems with varying representations. 
%which, when taken together, present a similar picture to the literature concerning two-agent systems.
%present a somewhat surprising picture.

Our approach follows the approach outlined in \cite{RV21} in that we consider the Nash equilibria as two separate conditions - one that corresponds to correct behavior under deviation (the $j$-Deviant Trace Condition) and one that corresponds to correct behavior when no deviations are observed (the Primary-Trace condition). Using this characterization we are able to prove a suite of new results and prove novel variants of a few older results that appeared in previous works under this unified framework. Taken together, they represent a complete characterization of the complexity of both problems for the three main types of automata-theoretic representations common to the literature on finite-horizon temporal logic.

\section{Background}

The background presented here largely follows \cite{RV21}. We assume familiarity with automata theory, as in \cite{sipser2006,Var96}.

\subsection{Games}

% \textbf{This section should be cut a lot as well}
In this section we provide some definitions related to two-agent games to provide a standard notation throughout this paper. The two agents are denoted agent $0$ and agent $1$.
\begin{definition} [Arena]

An \emph{arena} is a four tuple $A = (V, V_0, V_1, E)$ where $V$ is a finite set of vertices, $V_0$ and $V_1$ are disjoint subsets of $V$ with $V_0 \cup V_1 = V$ that represent the vertices that belong to agent $0$ and agent $1$ respectively, and $E \subseteq V \times V$ is a set of directed edges, i.e. $(v, v') \text{ } | \text{ } \in E$ if there is an edge from $v$ to $v'$.

Intuitively, the agent that owns a node decides which outgoing edge to follow.
Since $V = V_0 \cup V_1$, we omit $V$ and write $A =( V_0 , V_1, E)$.
\end{definition}

\begin{definition}[Play]
A \emph{play} in an arena $A$ is an infinite sequence $\rho = \rho_0 \rho_1 \rho_2 \ldots \in V^{\omega}$ such that $(\rho_n , \rho_{n+1}) \in E$ holds for all $n \in \mathbb{N}$. We say that $\rho$ starts at $\rho_0$
\end{definition}

%We now introduce a very broad definition for two-agent games.
\begin{definition}[Game]
A \emph{game} $G = (A,Win)$ consists of an arena $A$ with vertex set $V$ and a set of winning plays $Win \subseteq V^{\omega}$. A play $\rho$ is winning for agent $0$ if $\rho \in Win$, otherwise it is winning for agent $1$.

\end{definition}

Note that in this formulation of a game, reaching a state $v \in V$ with no outgoing transitions is always losing for agent $0$, as agent $0$ is the one that must ensure that $\rho$ is infinite ( a member of $V^{\omega}$).

 A game is thus defined by its set of winning plays, often called the \emph{winning condition}. One such widely used winning condition is the \emph{safety condition}.

\begin{definition}[Safety Games]
Let $A = (V, V_0, V_1, E)$ be an arena and $S \subseteq V$ be a subset of $A$'s vertices. Then, the \emph{safety condition} $Safety(S)$ is defined as  $Safety(S) = \{ \rho \in V^{\omega} \text{ } | \text{ } Occ(\rho) \subseteq S \}$,
where $Occ(\rho)$ denotes the subset of vertices that occur at least once in $\rho$. 
A game with the safety condition for a subset $S$ is a \emph{safety game} with the set $S$ of \emph{safe vertices}. Information about solving safety games, including notions of \emph{winning strategies} and \emph{winning sets} can be found here \cite{McNaughton1993InfiniteGP}.
\end{definition}

\subsection{Concurrent Games and iBGs}\label{iBGs}
A \emph{concurrent game structure} (CGS) is an 8-tuple 
$$( Prop, \Omega, (ACT_i)_{i \in \Omega}, S, \lambda, \tau, s_0 \in S, (A^i)_{i \in \Omega})$$ %inlining is just too ugly, I'll come back here as a last resort
where $Prop$ is a finite set of \emph{propositions}, $\Omega =\{ 0, \ldots k-1 \} $ is a finite set of \emph{agents}, $ACT_i$ is a set of \emph{actions}, where each $ACT_i$ is associated with Agent $i$, the set of \emph{decisions} is $D =  ACT_0 \times ACT_1 \ldots ACT_{k- 1}$,  $S$ is a set of \emph{states}, $\lambda: S \rightarrow 2^{Prop} $ is a \emph{labeling function} that associates each state with a set of propositions that are interpreted as true in that state, $\tau : S \times D \rightarrow S$ is a deterministic \emph{transition function} that takes a state and a decision as input and returns another state, $s_0 \in S$ is the  \emph{initial state}, and $A^i$ is a \emph{goal} specification for  Agent $i$ given in the form of an deterministic finite automaton (DFA), nondeterministic finite automaton (NFA), or alternating finite automaton (AFA).
\footnote{For automata-theoretic background, see \cite{Var96}}
We say that a finite word automaton accepts an infinite word $\omega$ if it accepts a finite prefix of $\omega$. In a CGS, Agent $i$ prefers plays in the game that satisfy $A^i$, that is, a play such that some finite prefix of the play is accepted by $A^i$. For a goal automaton we use the notation $A^i=\langle Q^i, q^i_0 , \Sigma, \delta^i,  F^i \rangle$, where $Q^i$ is the state space, $q^i_0 \in Q^i $ is the initial state, $\Sigma$ is the alphabet, $\delta^i$ is the transition function, and $F^i$ is the set of final states.

We now define \emph{iterated boolean games} (iBG), a restriction on the CGS formalism \cite{iBG}. We follow the formulation of \cite{RV21}, as we take the set of actions to be a finite alphabet rather than a set of truth assignments. An iBG is defined by applying the following restrictions to the CGS formalism. Agent $i$ `owns'' alphabet $\Sigma_i$. These $\Sigma_i$ are disjoint and each $\Sigma_i$ serves as the set of actions for Agent $i$-an action for agent $i$ consists of choosing a letter in $\Sigma_i$. The set of decisions is then $\Sigma = \bigtimes_{i=0}^{k-1} \Sigma_i$. The set of states is also $\Sigma$, and the labeling function is the identity function, i.e., $\lambda(s) = s$. Our use of the iBG model is motivated by presentation, as iBGs offer a simple model in which agent actions influence global states - considering general CGS models would not influence the forthcoming complexity-theoretic results. 
%\myv{Deleted}
%\sr{Okay}
%As a slight abuse of notation, we consider $\sigma \in \Sigma_i$ to take on the role of a ``proposition'' , and is true at a state $s$ if $\sigma$ appears in $s$, allowing us to generalize towards arbitrary alphabets. 
Finally, the transition function $\tau$ is simply the right projection $\tau(s,d) = d$. 

We now introduce the notion of a \emph{strategy} for Agent~$i$ in the general CGS formalism.
\begin{definition}[Strategy for Agent $i$]
A strategy for Agent $i$ is a function $\pi_i : S^* \rightarrow ACT_i$. Intuitively, this is a function that, given the observed history of the game (represented by an element of $S^*$), returns an action $act_i \in ACT_i$.
\end{definition}

Recalling that $\Omega = \{ 0, 1 \ldots k-1 \}$ represents the set of agents, we now introduce the notion of a \emph{strategy profile}.
\begin{definition}[Strategy Profile]\label{Strategy Profile}
Let $\Pi_i$ represent the set of strategies for agent $i$.  We define the set of strategy profiles $ \Pi = \bigtimes_{i \in \Omega} \Pi_i$ and denote a single strategy profile as $\pi$.
\end{definition}
A strategy profile can be naturally thought of as a function of type $\Sigma^* \rightarrow \Sigma$, and we will call any function with such type a \emph{global strategy}. Since a global strategy is deterministic, it yields a unique element of $S^{\omega}$, which we call a \emph{primary trace}.

\begin{definition}[Primary Trace resulting from a Global Strategy]\label{Primary Trace}
Given a global strategy $\pi:\Sigma^* \rightarrow \Sigma$, the primary trace of $\pi$ is the unique trace $t$ that satisfies
\begin{enumerate}
    \item $t[0] = \pi(\epsilon)$
    \item $t[i] = \pi( t[0], \ldots t[i-1])$
\end{enumerate}
We denote this trace as $t_{\pi}$.
\end{definition}
Given a trace $t \in S^\omega$, define the \emph{winning set} $W_t=\{i\in\Omega~:~ t\models A^i \}$ to be the set of agents whose DFA goals are satisfied by a finite prefix of the trace~$t$. The \emph{losing set} is then defined as $ \Omega / W_t$.

A common solution concept in game theory is the \emph{Nash equilibrium}, which we adapt to our iBG framework. In our framework, a Nash equilibrium is a strategy profile $\pi$ such that for each Agent~$i$, if $A^i$ is not satisfied on $t_{\pi}$, then a unilateral strategy deviation for Agent~$i$ does not result in a trace that satisfies $A^i$. 
 \begin{definition}[Nash Equilibrium]\cite{iBG}
Let $G$ be an iBG and $\pi = \langle \pi_0 , \pi_1 \ldots  \pi_{k-1}\rangle$ be a strategy profile. We denote $W_{\pi}=W_{t_\pi}$.  The profile $\pi$ is a \emph{Nash equilibrium} if for every $i \in \Omega / W_t$ we have that for each strategy profile of the form $\pi' = \langle\pi_0 , \pi_1 \ldots \pi'_i \ldots  \pi_{k-1}\rangle$, with $\pi^{'}_i \in \Pi_i$, it is the case that $i \in \Omega / W_{\pi '}$.
\end{definition} 
This definition provides an analogy for the Nash Equilibrium defined in \cite{Nash48} by capturing the same property - no agent can unilaterally deviate to improve its own payoff (moving from an unsatisfied goal to a satisfied goal). Agents in the set $W_{\pi}$ cannot have their payoff improved further, so we do not check their deviations. We say that $\pi$ is a $W$-NE iff $\pi$ is a Nash Equilibrium with $W_{\pi} = W$.

We have already defined the primary trace, which corresponds to the trace that results from no deviations to a profile $\pi$. Since we consider unilateral deviations from a single agent in our analysis, we define these traces as well.

\begin{definition}[$j$-Deviant-Trace from a Strategy Profile] Given a strategy profile $\pi$, a $j$-Deviant-Trace (w.r.t $\pi)$ is defined as follows.
For $\alpha \in \Sigma$,  we introduce the notation 
$\alpha[-j]$ to refer to $\alpha|_{\Sigma \setminus \Sigma_j}$ (that is, $\alpha$ with $\Sigma_j$ projected out). A trace $t=y_0,y_1,\ldots$ is $j$-deviant if
\begin{enumerate}
\item
$y_0=\varepsilon$
\item
$y_{i+1}=y_0,\ldots,y_i,\alpha$, where $\alpha\in \Sigma$ and 
$\alpha[-j]=\pi(y_i)[-j]$
\item $t$ is not the primary trace
\end{enumerate}
\end{definition}

Our characterization of the Nash equilibria is based on \cite{RV21}, in which the Nash Equilibrium condition was decomposed into the \emph{Primary-Trace} and \emph{j-Deviant-Trace} Condition, which we reintroduce here. For a fixed strategy profile $\pi$ we have:
\begin{enumerate}
\item {\sf Primary-Trace Condition}: The primary infinite trace $t_\pi$ defined by $\pi$ satisfies the goals $A^i$ for precisely  $i\in W$. 

\item {\sf $j$-Deviant-Trace Condition}:
Each $j$-\emph{deviant} trace $t=y_0,y_1, \ldots$ (w.r.t $\pi)$ for $j\not\in W$, does not satisfy the goal $A^j$.
\end{enumerate}

A strategy profile $\pi : \Sigma^* \rightarrow \Sigma$ is then a $W$-NE  iff it satisfies both the Primary-Trace Condition and the $j$-Deviant Trace Condition w.r.t $W$. In this definition we work with a general notion of the goal $A^j$ accepting a prefix, therefore this universally applies to DFA, NFA, and AFA goals.

The main insight of \cite{RV21} was to show that these two conditions could be reasoned about through automata-theoretic means \emph{separately} when considering DFA goals. In order to analyze the $j$-Deviant-Trace Condition for a single agent $j$ with goal $A^j$, a safety game $G_j$ with $V_0 = Q^j$ and $V_1=\{Q^j \times \Sigma\}$ and edge relation  $E_j$  is defined as follows:
\begin{enumerate}
\item 
$(q,\langle q,\alpha\rangle)) \in E_j$ for $q \in Q^j\setminus F^j$ and all $\alpha \in \Sigma$.
\item 
$(\langle q,\alpha\rangle, q') \in E_j$ for $q\in Q^j$ and $q'\in Q^j$, where $q'=\delta^j(q,\beta)$ for some $\beta\in \Sigma$ such that $\alpha[-j]=\beta[-j]$.
\end{enumerate} 
The winning set of agent $0$ in this game is denoted $Win_0(G_j)$; the winning set for agent $1$ is $Win_1(G_j)$.

In order to reason about the Primary Trace, we construct the deterministic B\"uchi automaton \cite{Var96} $A_W = \langle Q,q_0,\Sigma,\delta,F \rangle$ with \begin{enumerate}
\item
$Q=(\bigtimes_{j\in\Omega} Q^j) \times 2^\Omega$ 
\item
$q_0=\langle q^1_0,\ldots,q^n_0,W \rangle$
\item
$F=(\bigtimes_{j\in\Omega} Q^j) \times \{\emptyset\}$
\item
$\delta(\langle q_1,\ldots q_n,U\rangle,\alpha)=\langle q'_1,\ldots q'_n,V\rangle$ if $q'_j=\delta^i(q_j,\alpha)$, 
where $q'_j\not\in F^j$ for $j\not\in W$, and 
$V=U-\{i: q_i'\in F^i\}$.
\end{enumerate} The intuition is that a word in $\Sigma^{\omega}$ is accepted by $A_W$ iff only the goals in $W$ are satisfied on it. 

The winning sets for agent $0$ in the games $G_j$ are now used to refine the state space and transition function of $A_W$ to create the deterministic B\"uchi automaton $A'_W=(Q',q_0,\Sigma,\delta',F \cap Q')$, with $Q' = \bigtimes_{i \in W} Q^i \times \bigtimes_{j\in \Omega\setminus W} \{ Win_0(G_j) \cap Q^j \} \times 2^{\Omega}$, and $\delta'$ defined as follows:
$\delta'(q,\alpha)=\delta(q,\alpha)$ if, for all $j \not \in W$, we have that $(q[j], \alpha) \in Win_0(G_j)$; otherwise, $\delta'(q,\alpha)$ is undefined. It was then shown that 
\begin{theorem}\emph{\cite{RV21}}
For a given iBG $G$, a $W$-NE strategy exists in $G$ iff $A'_W$ is nonempty.
\end{theorem}%

Since the goal automata being considered have an input alphabet that is the cross product of $k$ other alphabets ($\Sigma = \Sigma_0 \times \Sigma_1 \ldots \Sigma_{k-1}$) , they can be seen as exponential constructions themselves. For this reason, we introduce \emph{bounded-channel automata}.% Note that if $|A|$ is a constant w.r.t $|\Omega| = k$, then such automata are polynomial-sized constructions instead.

\begin{definition}[Bounded-Channel Automaton]
Let $\Sigma = \Sigma_0 \times \Sigma_1 \ldots \Sigma_{k-1}$ and let $I \subset \{ 0 \ldots k-1 \}$ be a strict subset of agents.  A \textit{bounded-channel automaton} is an automaton with a transition function $\rho$ that satisfies the property that for all $\alpha,\beta \in \Sigma$ and states $q$ in the automaton, if $\alpha_I = \beta_I$, i.e $\alpha$ and $\beta$ agree on $\Sigma_i$ for every $i\in I$, then  $\rho(q,\alpha) = \rho(q,\beta)$, i.e. the transition function only considers $\Sigma_i$ for $i \in I$ on a state $q$ and a  symbol $\alpha \in \Sigma$.
\end{definition}
Intuitively, a bounded-channel automaton does not consider the actions of every agent through an element of $\Sigma$ but has an input alphabet $\Sigma_B = \bigtimes_{ i \in I} \Sigma_i$ where $I \subset \Omega$. Note that if $|I|$ is a constant w.r.t $|\Omega| = k$, then such automata have polynomial-sized alphabets. 
%This is due to the fact that it only needs to consider the settings of a constant number of the $\Sigma_i$, of which there are only a polynomial number of possible configurations.  Therefore, we can reason about succinct goal representations without incurring the blow-up of taking the cross product of all $\Sigma_i$. 
While the use of bounded-channel goal automata does not affect many of the complexity results in this paper, considering bounded channel automata allows us to reason about a more succinct input type that arguably better corresponds  to realistic situations.
%When possible, we employ the same approach here.   %This approach is used in this paper as well - in order to analyze the $W$-NE condition, we create automata that analyze the Primary-Trace Condition and automata that analyze the $j$-Deviant-Trace Condition, unifying them when possible. 
%\myv{Here you lost the reader.}
%\sr{I've tried to make this more clear? Without knowing the exact problem, I don't know if I addressed it or not.}
%This was done through the construction of a safety game for every deviating Agent $j$, denoted $G_j$, which corresponded to analyzing the $j$-Deviant-Trace Condition for a single Agent $j$. This allowed for the refinement of the B\"uchi word automaton $A_W$ that checked the Primary Trace Condition into a B\"uchi word automaton $A'_W$ that checked both conditions by restricting the states and transitions of $A_W$ based on the winning regions of each $G_j$\cite{RV21}. When possible, we also utilize this same approach.

\section{Realizability}
In this section we study the \emph{realizability problem} (referred to as the \emph{nonemptiness problem} in \cite{iBG,RV21}) in which we are are given an iBG $G$ \emph{and} a set $W \subseteq \Omega$ of agents and we wish to decide if a Nash equilibria strategy profile exists in which only the agents in $W$ have their goals satisfied. In \cite{RV21}, this problem was proven to be PSPACE-complete for DFA goals, using the safety games $G_j$ and B\"uchi automaton $A'_W$, as described in section~\ref{iBGs}. We now analyze the complexity for NFA and AFA goals.

\subsection{NFA Goals}
Assume the input goal automata are NFAs. By determinizing each goal automaton, we can readily apply the procedure from \cite{RV21}. Constructing the DFA goal automaton $A^j=\langle Q^j, q^j_0 , \Sigma, \delta^j,  F^j \rangle$ from the NFA goal input involves a worst-case exponential blowup in the number of states, with no blow-up in the size of the alphabet $\Sigma$. Therefore, the state space of each safety game $G_j$, given by $ Q^j \cup \{Q^j \times \Sigma \} $ is overall exponential in the size of the goal NFAs. Safety games can be solved in time linear in the size of the game \cite{McNaughton1993InfiniteGP}, so each relevant $G_j$ can be analyzed in EXPTIME.

The automaton $A'_W$ from \cite{RV21} has a state space $Q'$ that is upper bounded by the size of the cross product of all DFA goals and $2^{\Omega}$. Each DFA goal has a state space that is exponential in the size of the NFA, so the  product of all DFAs is still singly exponential with respect to the NFA goals. Furthermore, $2^{\Omega}$ is also singly exponential, implying that $A'_W$ is singly exponential in size with respect to the input. Testing a B\"uchi automaton for nonemptiness can be done in NLOGSPACE \cite{VW94}, meaning that $A'_W$ can be checked for nonemptiness in PSPACE. Since the safety games are in EXPTIME in the worst case, the overall complexity of this method is in EXPTIME. These results still hold when considering bounded-channel goals, as they are still polynomial in the size of the input. 

\begin{theorem}
The realizability problem with NFA goals can be solved in EXPTIME.
\end{theorem}

Now, we show that the realizability problem with NFA goals is EXPTIME-hard through a polynomial time reduction from the following EXPTIME-hard problem : Given an alternating Turing machine $M$ and a number $n \in \mathbb{N}$ written in unary, does $M$ accept the empty tape using at most $n$ cells? This problem is a canonical hard problem for the set of languages recognized by alternating polynomial-space Turing machines, which is equivalent to the set of languages recognized by deterministic exponential-time Turing machines (i.e., EXPTIME) \cite{CKS}.

An alternating Turing machine is a generalization of nondeterministic Turing machines. In a nondeterministic Turing machine, it is possible for a machine in state $\alpha$ to transition to multiple states; with out loss of generality we can assume there are at most two possible successor states -- $\beta_0$ and $\beta_1$, which we call the \emph{left} and \emph{right} successors, respectively.  The computation from $\alpha$ is accepting iff the computation from at least one $\beta_i$ is accepting. This means the machine ``chooses'' the best option and proceeds from there. Alternating Turing machines classify states as either existential $(\vee)$ or universal $(\wedge)$. In an existential state the machine is allowed to pick the best transition possible for acceptance, analogous to nondeterminism. In a universal state \emph{all} successor states must lead to acceptance. This dynamic is usually modeled by a two-agent game in which a second agent is antagonistic to the machine's `goal' of acceptance makes the \emph{worst} possible transitions for acceptance at universal states. We also assume that $M$ has unique accepting and rejecting states. Acceptance in an alternating Turing machine is then characterized by an accepting computation \emph{tree} as opposed to a single accepting run of a nondeterministic machine. Such a tree demonstrates the computation run of the machine for every choice of successor state made at universal states. Since $M$ is space-$n$ bounded, we can assume without loss of generality that each computation of $M$ terminates in exponential time in an accepting or rejecting state.

We now create a two-agent (labeled $0$ and $1$) iBG $G$ with NFA goals such that a $\{0\}$-NE exists in $G$ iff the alternating polynomial space Turing machine $M$ accepts the empty tape using at most space $n$. We start by defining Agent 0's alphabet $\Sigma_0 = \Delta \cup \{\Delta \times R \} \cup \{\#\}$, where $\Delta$ is the alphabet of $M$, $R$ is the state set of $M$, and $\#$ is a new symbol that does not appear in $\Delta$. Agent $1$'s alphabet $\Sigma_1$ is given by $\{0,1\}$, which are symbols that do not appear in $\Delta$ by assumption.

In order to analyze the computations of $M$, we introduce the notion of an ID of a Turing machine, which is a string that represents the content of the tape at a particular time step in the run of $M$. Such an ID includes
\begin{itemize}
\item The complete content of the tape from left to right.
\item The position and state of the head of $M$. As a matter of notation, if the head is on cell~$i$, then the character corresponding to the content of cell~$i$ is a pair consisting of the element of $\Delta$ on the tape \emph{and} an element of $R$ representing the state of the machine.
\item A $\#$ as the first symbol.
\end{itemize}
For example, an ID may read $\#12\bot 3 \langle 0,q \rangle 22$, meaning that the contents of the tape read $12\bot 3 0 22$ with $\bot$ representing an empty cell (we assume WLOG that this is a special symbol in $\Delta$), and the pair $\langle 0,q \rangle $ denotes that the head is currently reading the cell with $0$ written in it in and the machine in state $q$.

Intuitively, Agent $0$ generates IDs of the machine $M$ and has a goal expressing $M$'s accepting the empty tape. Agent $1$ takes on the role of the antagonist in the alternating Turing machine $M$ and chooses the successors at universal states by specifying either $0$ or $1$ to differentiate between the two possible options. Agent 1's goal is to witness either $M$ rejecting the empty tape \emph{or} to witness some sort of computation error in the IDs generated by agent $0$. Since the machine $M$ is restricted to using at most $n$ cells and each ID represents the total state of the machine, each ID is $n+1$ symbols long. Agent $0$ thus describes a single ID over the course of $n+1$ successive time steps in $G$ by outputting the symbols of the ID in left to right order. When agent $0$ outputs a character in $\{\Delta \times R \}$ specifying the state of the machine, Agent $1$'s decision of $0$ or $1$ becomes relevant if the state $r \in R$ is universal, as in this case it is Agent $1$ who specifies the successor state of $r$.

We now describe the NFA goals of each agent. Agent $0$'s goal checks that the computation ends in an accepting state. Thus, Agent $0$ wishes 
%to see the correct starting ID specified by $\langle \bot, q_0 \rangle \bot^{n-1}\#$, and then 
to eventually see a symbol $\langle a, q_A \rangle$, where $q_A$ is the unique accepting state, for some $a \in \Delta$. A trace without ``errors'' that satisfies Agent $0$'s goals is a correct branch of an accepting computation tree of $M$.

%It also checks to verify that $M$ never moves out of bounds by $M$ moves out of bounds by observing the first non $\#$ cell and the last cell in an ID and seeing if the transition functions specifies a move to the left or right respectively (we include this specification in this goal as well in order to make agent $0$'s goal mutually exclusive to agent $1$'s). If such a computation can be demonstrated with no errors for agent $1$ to accept on, then $M$ accepts the empty tape using at most $n$ cells.

Agent $1$'s goal is to show that the computation described the the ID sequence generated by Agent $0$ contains an error or is rejecting. Thus, the goal is an NFA $A^1$ that represents \emph{all} possible ways this ID sequence is not a valid acceptance computation of $M$ on the empty tape using at most $n$ cells. This consists of several disjoint checks, which can all be expressed by an NFA of size polynomial in $n$ and $M$.
%which we specify through the use of regular expressions. For the purpose of presentation, we express these regular expressions in terms of $\Sigma_0$ even though the structure of the game is defined using $\Sigma$. This is due to the fact that $\Sigma^*_0$ readily corresponds to IDs, and $\Sigma_1$ is only taken into account sometimes to determine state-successors of universal states. 
\begin{itemize}
    %\item The word over $\{0,1\}$ generated by Agent $1$ must be a sequence of blocks of the form $0^{n+1}$ or $1^{n+1}$. This says that each ID generated by Agent $0$ is labeled by Agent $1$ asa $0$-ID or $1$-ID.
    \item The reject state $q_R$ is seen. 
    %This is specified by the regular expressions of the form $\Sigma_0^* \langle w, q_R \rangle$ for any $w \in \Delta$ and $q_R$ the unique rejecting state. If the machine does not reject, it is still possible for agent $1$ to catch a computational error as described by the following conditions.
    \item The initial ID generated by Agent 0 is wrong, and does not represent the empty tape with the head at the left in the starting state of $M$.
    \item The trace generated by Agent 0 is not a sequence of of blocks of size $n+1$, each starting with the symbol $\#$, which does not appear anywhere else.
    %The machine $M$ starts wrong - i.e. the tape is not empty at the start. This is represented by the negation of $\# \langle \bot, q_0 \rangle \bot^{n-1}$, where  $q_0$ is the initial state which we assume WLOG starts all the way to the left.
    %through regular expressions of the form $\Sigma_0^* a b c \Sigma_0^n d \Sigma_0^* $ where $d$ does \emph{not} follow from the previously seen characters $a,b,$ and $c$ by the transition function of $M$. Note that this also includes cases in which the specification of agent $1$ through $\Sigma_1$ is not respected in the case of universal transitions - though we do not include $\Sigma_1$ in the regular expression for the sake of presentation. 
    %Since we are considering consecutive blocks of three characters, the number of regular expressions is $O(n \times (|\Sigma_0| \times |\Sigma_1|)^4)$ (corresponding to all possible configurations of $4$ characters), which is polynomial w.r.t to the input $M$. Note that $|\Sigma_1| = 2$.
    \item Agent 0 output a malformed ID, where the head is positioned at two distinct cells.
    %\item The machine outputs a malformed ID - i.e. one that is not $n+1$ characters long. This can by observed by with regular expressions of the form $\Sigma_0^* \# \Sigma^i_0 \#$ for all $i < n$ and a regular expression of the form   $\Sigma_0^* \# (\Sigma_0 \setminus \#)^{n+1}$ to capture longer IDs.
    %\item The machine outputs an ID with two separate states. This can be recognized with regular expressions of the form $$\Sigma^*_0\# (\Sigma_0 \setminus \#)^* a (\Sigma_0 \setminus \#)^* b (\Sigma_0 \setminus \#)^* \#$$ where both $a,b \in \Delta \times R$. 
    %\item $M$ moves out of bounds - This can be accomplished by observing the first non $\#$ cell and the last cell in an ID and seeing if the transition functions specifies a move to the left or right respectively. As an example, it consist of expressions of the form $\Sigma_0^* \# \langle w, q \rangle$ where $w \in \Delta$ and $q\in R$ is a state that moves left upon reading $w$. Note that there are only polynomial many invalid configurations to record, as they are a subset of $\Delta \times R$.
    \item The machine $M$ transitions incorrectly - i.e. the transition function of $M$ is not respected. This is characterized by the local nature of Turing machines, whereby the content of three cells -- $\langle c_{i_1},c_i,c_{i+1}\rangle$ in an ID constrains the content of these cells in the successor ID, in terms of the new state, the new content of cell where the machine head was positioned, and the new position of the head. If the state is existential, then either one of the two possible transitions has to be taken. But if a state $q$ is universal, then the symbol out output by Agent 1, in the round when Agent 0 output the symbols $\langle q,a\rangle$, for some $a\in\Delta$. If that symbol is $0$, then the correct transition is the left one, and if that symbol is $1$, then the correct transition is the right one.
\end{itemize}

\begin{lemma}
$M$ accepts the empty tape in at most $n$ cells iff a $\{0\}-NE$ exists in $G$.
\end{lemma}

\begin{proof}
$(\rightarrow)$ Suppose that agent $M$ accepts the empty tape using at most $n$ cells. This means that there is a valid tree of ID sequences that start from the proper initial configuration and end in an ID with the unique accepting state no matter what choice is made at the universal states. This tree of ID sequences is used to create an agent $0$ strategy $\pi_0$ in $G$. 

The strategy $\pi_0$ is given exactly by this ID tree, but takes $n+1$ time steps to output a single ID. Observing Agent $1$'s previous choices for universal state transitions informs agent $0$ on which branch of the ID sequence tree to follow. Since all branches in the ID tree sequence are valid, accepting computation runs, agent $1$ is not be able to find an error no matter what choices he makes at the universal states, nor is he able to witness the rejecting state. Therefore, $\pi_0$ ensures that Agent $0$'s goal is satisfied on the primary trace of $\langle \pi_0, \pi_1 \rangle$ for an arbitrary Agent $1$'s strategy $\pi_1$ and and, furthermore, that there is no $\pi'_1$ such that $\langle \pi_0, \pi'_1 \rangle$ satisfies Agent $1$'s goals. Therefore, $\langle \pi_0, \pi_1 \rangle$ is a $\{0\}$-NE in $G$ given the choice of \emph{an arbitrary} Agent $1$ strategy $\pi_1$.

$(\leftarrow)$
Suppose that a $\{0\}$-NE exists in $G$. This means that there exists an Agent-$0$ strategy $\pi_0$ and an Agent-$1$ strategy $\pi_1$ such that the primary trace of $\langle \pi_0, \pi_1 \rangle$ satisfies Agent $0$'s goal but not Agent $1$'s goal. Furthermore, for every other Agent $1$ strategy $\pi'_1$,  $\langle \pi_0, \pi'_1 \rangle$ does not satisfy Agent $1$'s goal. By a dual logic to the argument shown for the $\rightarrow$ direction, we can use this $\pi_0$ to create a tree of ID sequences that all represent accepting computations of $M$. Since no branch of these tree reaches a rejecting state, as that would satisfy Agent $1$'s goal, every branch must reach an accept state.
%Note that since Agent $1$'s goal is mutually exclusive to agent $0$'s, each one of these branches is accepting even though the $\{0\}$-NE specification does not stipulate that agent $0$'s goal hold on an agent $1$-deviant trace. 
\end{proof}

Note that all constructions in $G$ were made in polynomial time. Agent $1$'s goal consists of the union of polynomially many polynomial-size NFAs, and Agent $0$'s goal is even simpler.
%a simpler union of three polynomial sized NFAs. 
Furthermore, $\Sigma_0$ and $\Sigma_1$ are clearly polynomial w.r.t $M$.  As such, $G$ can be constructed from $M$ and $n$ in polynomial time, giving us the desired polynomial-time reduction from an EXPTIME-hard problem to our NFA realizability problem. We further note that since a fixed number of agents (two) were considered, this reduction also holds for the bounded-channel automata case. Combining this lower bound with previously established EXPTIME upper bound yields 

\begin{theorem}
The realizability problem with NFA goals is EXPTIME-complete.
\end{theorem}

%We conjecture this bound to be tight, via a reduction from the acceptance problem of alternating polynomial space Turing machines.

\subsection{AFA Goals}
We now analyze the case of the realizability problem when the agents' goals are specified by AFAs. Since AFAs are linear (in number of states) in the size of equivalent finite-trace temporal specifications such as $LTL_f$ or $LDL_F$ \cite{GV13}, we note that a similar problem of deciding whether any $W$-NE exists was given a 2EXPTIME upper bound, but no lower bound, in \cite{GPW17}. Here we focus on the a single agent set $W$ and provide a tight 2EXPTIME bound.
%we show the method in \cite{RV21} yields the same upper bound for the single $W$-NE problem for AFA inputs (and if all $W$ need to be tested as in \cite{GPW17} this represents a singly exponential blowup, maintaining 2EXPTIME) with a decision procedure that is arguably easier to understand while proving a matching lower bound. 

Constructing the DFA goal automata $A^j$ from the AFA goals involves a doubly exponential worst-case blowup in the number of states \cite{GV13}, with no blow-up in the size of the alphabet $\Sigma$ - this holds for bounded-channel automata as well. The analysis now largely follows the NFA-goals case. The size of the safety game $G_j$ is now doubly exponential in the size of the input due to the presence of $Q^j$, and so each $G_j$ can be solved in 2EXPTIME. Meanwhile $A'_W$ consists of the cross product of the doubly exponential $Q^j$s and the singly exponential $2^{\Omega}$, so it is doubly exponential overall. Following the same logic as before yields a 2EXPTIME upper bound.

\begin{theorem}
The realizability problem with AFA goals can be solved in 2EXPTIME.
\end{theorem}

This result agrees with the result in \cite{GPW17} (where $W$ is not part of the input). We now extend the analysis by providing a matching lower bound, proving the problem to be 2EXPTIME-hard by reducing from the 2EXPTIME-complete problem of $LTL_f$ realizability, noting that there is a linear-time conversion from $LTL_f$ formulas to equivalent AFA for a fixed-size alphabet \cite{GV13}. We note that the 2EXPTIME lower bound for $LTL_f$ realizability holds already for fixed alphabet goals, as in \cite{Ros92}.

The $LTL_f$ \emph{realizability problem} \cite{GV15} takes as input an $LTL_f$ formula $\phi$ along with a partition of the variables $V$ in $\phi$ into two sets $X$ and $Y$. The problem asks whether an agent (Agent 0) that takes control of the variables in $X$ can always ensure a trace satisfying $\phi$ with an antagonistic agent (Agent 1) setting the variables in $Y$. At each time step the agents set their variables, thus producing an infinite trace over $2^{X\cup Y}$. As before, $\phi$ is satisfied by an infinite trace if it is satisfied by some finite prefix of that infinite trace. The interaction is naturally modeled as a game between the two agents, which is called the $\phi$-\emph{realizability game}. There are several variations of these games, we consider the variation where Agent 0 and Agent 1 move concurrently, assigning values to the $X$ and $Y$ variables, respectively \cite{GV15}.

%The description above is intentionally informal due to the fact that there are many models of the aforementioned game. For our purposes we employ the model in which the agent setting $Y$ assigns their variables first, which is known to be 2EXPTIME-complete \cite{GV13}.

Given an instance of the $LTL_f$ realizability problem with goal $\phi$, we construct an iBG $G_\phi$ with two agents. Agent 0 is given the goal $\phi$ expressed as AFA and Agent 1 is given an empty goal, i.e., an AFA that accepts the empty language. Let $\Sigma_0 = 2^X$, $\Sigma_1 = 2^Y$, and the set $W$ be the empty set. Since we assume that the temporal goal has a bounded alphabet, the translation to AFAs is linear in number of states.
%While it may seem that an exponential blowup is needed in order to represent $\Sigma = \Sigma_0 \times \Sigma_1$, we note that the problem of $LTL_f$ realizability is still 2EXPTIME-complete even when bounded alphabets are considered \cite{Ros92}, so the size of $\Sigma$ can be treated as a constant. This implies that the overall blowup from $LTL_f$ realizability to an AFA game is polynomial.

\begin{theorem}
Given an $LTL_f$ formula $\phi$, Agent $0$ wins the $\phi$-realizability game iff no $\emptyset$-NE exists in $G_\phi$.
\end{theorem}
\begin{proof}
($\rightarrow$)
Assume that Agent $0$ wins the $\phi$-realizability game. Then, there exists a strategy $\pi'_0 : (2^Y)^* \rightarrow 2^X $ that ensures that the formula $\phi$ is eventually satisfied given an arbitrary Agent 1 strategy $\pi_1 : (2^X)^* \rightarrow 2^Y$.
Therefore, there can not be $\emptyset$-NE in $G$. Suppose to the contrary that the profile $\langle \pi_0,\pi_1\rangle$ is an $\emptyset$-NE, which means that the primary trace does not satisfy $\phi$. Agent 0 can now deviate from this profile and follow the strategy $\pi'_0$, so now Agent 0 and Agent 1 are following the profile $\langle \pi'_0,\pi_1\rangle$. Since $\pi'_0$ is a winning strategy in the $\phi$-realizability game, Agent 0 is able to force satisfaction of $\phi$, so $\langle \pi_0,\pi_1\rangle$ is not an $\emptyset$-NE.

%\myv{This argument is not clear}
%Regardless of the strategy followed by Agent 1, by following the strategy $\sigma_0$, Agent $0$ is able to deviate from the start of the game to produce a trace that satisfies $\phi$ for sure. Once agent $0$ deviates, agent $1$ is required to move first in the resulting safety game, meaning that $\sigma_0$ will be winning in this game. Since a deviation from agent $0$ is always possible in this case, there can never be an $\emptyset$-NE.

($\leftarrow$)
Assume to the contrary that there is an $\emptyset$-NE in $G_{\phi}$. We show that it implies that Agent $1$ wins the $\phi$-realizability game.
Let $\langle \pi_0, \pi_1 \rangle$ be the strategy profile for the $\emptyset$-NE in $G_{\phi}$. This means that $\phi$ is not satisfied in the primary trace, and, furthermore, for every strategy profile $\langle \pi'_0, \pi_1 \rangle$ the primary trace does not satisfy $\phi$. This means that $\pi_1$ is winning strategy for Agent 1 in the $\phi$-realizability game.

%for the  exists in $G$. By definition, this means that for any other agent $0$ strategy $\sigma'_0$ the profile  $\langle \sigma'_0, \sigma_1 \rangle$ does not result in a trace that satisfies $\phi$. Since Agent $1$'s goal is always unsatisfiable, the same statement holds for Agent $1$.

%In particular, this means that should Agent $0$ deviate from time step $0$, there is a strategy $\sigma_1$ that wins for sure in the resulting safety game. By using the strategy $\sigma_1$ in the synthesis game, Agent $1$ will be able to keep Agent $0$ from satisfying $\phi$. If this were not the case, then agent $0$ would be able deviate successfully in $G$, contradicting the assumption that  $\langle \sigma_0, \sigma_1 \rangle$ is an $\emptyset$-NE. Therefore, Agent $1$ has a winning strategy in the synthesis game, concluding the proof.
\end{proof}

Since $LTL_f$-realizability is known to be 2EXPTIME-complete \cite{GV15}, we get:
\begin{theorem}
The realizability problem with AFA goals is 2EXPTIME-complete.
\end{theorem}
Since this lower bound was shown for two-agent games, it holds for the bounded-channel case as well.

\section{Verification}
We now address the verification problem in which we are given an iBG $G$, a set $W \subseteq \Omega$ of agents, and a strategy profile $\pi = \langle \pi_0 \ldots \pi_{k-1} \rangle$, where $k = |\Omega|$.  We are given the strategy profile in terms of the individual strategies: Each $\pi_i = \langle S^i, s^i_0, \Sigma, \Sigma_i, \rho^i, \gamma^i \rangle$ is a Moore machine \cite{sipser2006} that represents a function of type $\Sigma^* \rightarrow \Sigma_i$, where $S^i$ is the set of states, $s^i_0 \in S^i$ is the initial state, $\Sigma_i$ is the alphabet controlled by agent~$i$, 
$\rho^i: S^i\times\Sigma\rightarrow S^i$ is the transition function, and $\gamma^i: S^i \rightarrow \Sigma^i$ is the output function. 
$\Sigma=\bigtimes_{i\in\Omega}\Sigma_i$ 
is the common alphabet of the goals in $G$.
The \emph{verification problem} takes as inputs $G$, $W$, and $\pi$ and outputs whether $\pi$ is a $W$-NE in $G$.

We first construct a Moore machine for $\pi$ from $\pi_0 \ldots \pi_{k-1}$ using the standard product construction:   $\pi= \langle S, s_0, \Sigma, \Sigma, \rho, \gamma \rangle$, where 
\begin{enumerate}
    \item $S = \bigtimes_{i \in \Omega} S^i$.
    \item $s_0=\langle s^0_0,\ldots,s^{k-1}_0\rangle$.
    \item $\Sigma$ is both the input and output alphabet in $\pi$.
    \item The transition function $\rho$ is defined component-wise. 
    For $\overline{t}=\langle t_0,\ldots,t_{k-1}\rangle \in S$ and $\alpha \in \Sigma$, we have $\rho(\overline{t},\alpha) = \langle t'_0,\ldots,t'_{k-1}\rangle$, where $t'_i=\rho^i(t_i,\alpha)$, for 
    each $i \in \Omega$.
    \item The output function $\gamma$ is defined component-wise. For $\overline{t}=\langle t_0,\ldots,t_{k_1}\rangle \in S$, we have $\gamma(\overline{t})=\langle \sigma_0,\ldots,\sigma_{k-1}\rangle$, where $\sigma_i=\gamma^i(t_i)$,  for each $i \in \Omega$.
    \end{enumerate}
Note that the size of $\pi$ is exponential in the sizes of $\pi_0 \ldots \pi_{k-1}$.
%Note that there is an exponential blow-up in constructing $\pi$.

We now define the state outcome of running $\pi$ on input words in $\Sigma^*$, inductively:
\begin{itemize}
    \item $\pi(\varepsilon)=s_0$, and
    \item $\pi(w\alpha)=\rho(\pi(w),\alpha)$, for $w\in\Sigma^*$ and $\alpha \in\Sigma$
\end{itemize}

% We now define the state outcome of running $\pi$ on input words in $\Sigma^*$, inductively:
% \begin{itemize}
%     \item $\pi(\varepsilon)=s_0$, and
%     \item $\pi(wa)=\rho(\pi(w),a)$, for $w\in\Sigma^*$ and $a\in\Sigma$
% \end{itemize}
The output of running $\pi$ on a word $w \in \Sigma^*$ is then $\gamma(\pi(w))$. In this way,  we get that $\pi$ represents a function of type $\Sigma^* \rightarrow \Sigma$ ,i.e. a global strategy  (see Definition \ref{Strategy Profile}), which yields a primary trace (Definition \ref{Primary Trace}).

We can define bounded-channel Moore Machines for single-agent strategies in the same way we have defined bounded-channel goal automata. The construction is the same - instead of considering input alphabet $\Sigma$, we consider a restricted version that only considers some subset of agents. This allows us to consider single-agent strategies that have a succinct representation just as we considered goals with succinct representations.
We now proceed to analyze the complexity of verification with respect to DFA, NFA, and AFA goals.

\subsection{DFA Goals}
As described in Section~\ref{iBGs}, the existence of a $W$-NE in an iBG $G$ with DFA goals can be analyzed through solving safety games $G_j$ and testing a B\"uchi word automaton $A'_W$ for nonemptiness. Note, however, that $A'_W$ had a state space that consisted of the cross product of all DFA goals. Since here we have a strategy profile specified explicitly by the Moore machine $\pi = \langle \pi_0 \ldots \pi_{k-1} \rangle$, the Primary-Trace and $j$-Deviant-Trace Conditions can be checked separately for each agent, avoiding a cross-product construction. We begin our analysis by considering agents in $W$.
%Here, we follow the same constructions closely while modifying them to take into account the Moore machine $\pi$ that represents the strategy profile to check. The key idea is that the strategy is provided by $\pi$, so we have to incorporate $\pi$ in the constructions of the safety games $G_j$ and the B\"uchi automaton $A'_W$. We now move on to the analyzing the Primary Trace and Deviant Trace conditions for agents $j \not \in W$. As before, we first construct a safety game to characterize the set of states from which deviation is possible.
%\documentclass[sigconf]{aamas}

\paragraph{Checking Agents in $W$}
As mentioned before, we no longer need to create an automaton from the cross product of all DFA goals $A^i$ to form $A_{W,\pi}$, which checks the primary trace for \emph{all} agents at once. Since the primary trace of $\pi$ is uniquely determined (Definition 2.7),  we can check whether this trace satisfies the goal $A^i$ for each agent $i\in W$.

For each agent $i \in W$, we construct a DFA $A^i \times \pi$ as the product of the goal $A^i=(Q^i, q^i_0, \Sigma, \delta^i,  F^i)$ and $\pi$. In detail, $A^i\times\pi= (Q^i \times S, \langle q_0^i, s_0 \rangle, \emptyset, \tau^i, F^i \times S)$. Note that the alphabet of this automaton is empty, so transitions are defined between states. For $q\in Q^i$ and $s\in S$, We have $\tau^i(\langle q,s\rangle)=\langle q',s'\rangle$, where  $q'=\delta^i(q,\gamma(s))$ and $s'=\rho(s,\gamma(s))$. Satisfaction of $A^i$ on the primary trace now corresponds to nonemptiness of this product automaton (as the transition function $\tau^i$ simulates the run of $A^i$ on the primary trace of $\pi$), which means that a state $\langle f,s \rangle$ with $f \in F^i$ is reachable from $\langle q^i_0,s_0\rangle$.  This implies that a prefix of the primary trace is accepted by $A^i$. Note that the state space of $A^i \times \pi$ is of exponential size, since the state space of $\pi$ is of exponential size.  Nonemptiness in a DFA with an exponential state space can be decided in NPSPACE=PSPACE.

If $A^i\times\pi$ is empty for some $i\in W$. then $\pi$ is not a $W$-NE as the goal goal $A^i$ for an agent~$i$ is not satisfied on the primary trace. We refer to these nonemptiness queries of $A^i\times \pi$ as the $i$-queries.

\paragraph{Safety Game for Deviating Agents}
We now move on to the analyzing the Primary Trace and Deviant Trace conditions for agents $j \not \in W$. As before, we first construct a safety game to characterize the set of states from which successful deviations are possible.
We adapt the safety game $G_j$ to take in account the fact that we wish to check if $\pi$ is a $W$-NE.  Formally, we construct the safety  game $G_{\pi,j} = (Q^j\times S, Q^j \times S\times \Sigma, E_{\pi,j})$. Agent 0 owns $Q^j\times S$ and agent 1 owns $Q^j\times S\times \Sigma$. The edge relation $E_{\pi,j}$ is defined as follows:
\begin{enumerate}
\item 
$(\langle q,s \rangle, \langle q,s,\alpha\rangle)) \in E_{\pi,j}$ for $q \in Q^j\setminus F^j$, $s\in S$, and $\alpha =\gamma(s)$.
\item 
$(\langle q,s,\alpha\rangle, \langle q',s'\rangle) \in E_{\pi,j}$ for $q,q'\in Q^j$ and  $s,s'\in S$, where $q'=\delta^j(q,\beta)$ 
and $s'=\rho(s,\beta)$, for some $\beta\in \Sigma$ such that $\alpha[-j]=\beta[-j]$.
\end{enumerate}
As in $G_j$, if $q\in F^j$, then $\langle q,s\rangle$ has no successor node, and agent 0 is stuck and loses the game. Since $G_{\pi,j}$ is a safety game, agent 0's goal is to avoid states in $F^j$ and not get stuck. Unlike in $G_j$, however, agent 0 has no ``discretion'' in $G_{\pi,j}$; the move in state $(\langle q,s \rangle$ must be to $\langle q,s,\alpha\rangle)$, where $\alpha =\gamma(s)$.
Intuitively, we check whether agent $0$ can win this game while sticking to the strategy profile $\pi$. By keeping track of the state $s\in S$ of $\pi$, agent $0$ must move in accordance with $\gamma(s)$. Therefore, solving the safety game $G_{\pi,j}$ amounts to a reachability query; agent 1 wins if she can reach a state $\langle q,s\rangle$, with $q\in F^j$. Because graph reachability is in NLOGSPACE and the size of the game $G_{\pi,j}$ is exponential in the input due to the exponential state space $S$,  the game can be solved in NPSPACE=PSPACE. We denote the set of winning states for agent $0$ as $Win_0(G_{\pi,j})$; $Win_1(G_{\pi,j})$ is the set of winning states for agent $1$.
Note that, in particular, $F^j\times S \subseteq Win_1(G_{\pi,j})$.

\paragraph{Checking the Agents in $\Omega \setminus W$}\

For an agent $j \not \in W$, we construct a  DFA that checks that the goal $A^j$ is not satisfied  on the primary trace or on a deviant trace. In detail, $A^j \times \pi =( Q^j \times S, \langle q_0^j, s_0 \rangle, \emptyset, \tau^j, %F^j\cup maybe just include how it could contain triples for WIn_1
Win_1(G_{\pi,j}) )$ is defined in exactly the same way as the automaton $A^i\times\pi$ for $i\in W$  with the exception of the set of final states. We show below that if $A^j \times \pi$ is nonempty, then either the Primary Trace Condition or the Deviant Trace conditions is violated for Agent~$j$.

So, we must make sure that no state in $Win_1(G_{\pi,j})$ is reachable in $A^j \times \pi$ from  $\langle q^j_0,s_0 \rangle$. This is equivalent to the automaton being empty, so we have another nonemptiness query but now one that should fail. A path from $\langle q^j_0,s_0 \rangle$ to a state in $Win_1(G_{\pi,j})$ corresponds to either acceptance of $A^j$ on the primary trace or
a violation of the $j$-Deviant-Trace Condition , both of which contradicts $\pi$ being a $W$-NE. We refer to these nonemptiness queries as the $j$-queries (a reference to $j \not \in W$), and we note that they can be decided in NPSPACE=PSPACE by the exact same logic as $i$-queries. We now prove the correctness of checking the $i$-queries for the agents $i \in W$ and the $j$-queries for the agents $j \not \in W$.

\begin{theorem}\label{DFAmethod}
Given an iBG $G$ with DFA goal inputs, a strategy profile $\pi$ is a $W$-NE iff the $i$-queries succeed and the $j$-queries fail.
\end{theorem}

\begin{proof}
$(\rightarrow)$. Assume that $\pi$ is a $W$-NE. Therefore, it satisfies both the Primary-Trace Condition and the $j$-Deviant-Trace Condition. Since $\pi$ satisfies the Primary-Trace Condition, there is a path from $\langle q^i_0,s_0\rangle$ to $F^i\times S$ in $A^i\times \pi$, so
we have that the $i$-queries are successful.

Suppose now, for contradiction, that some $j$-query succeeds. This means that there is a run of $A^j$ on the primary trace that enters a state $\langle q^j,s\rangle$ in $Win_1(G_{\pi,j})$. There are now two cases:
\begin{enumerate}
    \item If $q^j \in F^j$, then $A^j$ has just accepted on the primary trace, contradicting the assumption that $\pi$ was a $W$-NE.
    \item Otherwise, Agent $j$ now has a winning strategy in $G_{\pi,j}$ from state $\langle q^j,s \rangle$. By following this strategy, the agent is able to reach a state in $F^j\times S$, which means that Agent $j$ constructed a $j$-Deviant-Trace that satisfies $A^j$, violating the $j$-Deviant-Trace Condition of $\pi$ and contradicting the assumption that $\pi$ was a $W$-NE.
\end{enumerate}
Therefore, we have that the $i$-queries must succeed and the $j$-queries must fail.

$(\leftarrow)$ Assume now that the $i$-queries succeed and the $j$-queries fail. We show that $\pi$ satisfies the Primary-Trace Condition and the $j$-Deviant-Trace Condition.

For the Primary-Trace Condition, note that goals $A^i$, for $i\in W$, accept on the primary trace of $\pi$, since the $i$-queries succeeded. Furthermore, the goals $A^j$, for $j\not\in W$,  cannot accept on the primary trace of $\pi$, as this would correspond to a path in $A^j\times\pi$ from $\langle q^j_0,s_0\rangle$ to $F^j\times S \subseteq Win_1(G_{\pi,j})$ in $A^j \times \pi$. No such path exists since the $j$-queries failed.

For the $j$-Deviant-Trace Condition, we only need to study the $j$-queries, $j\not\in W$. Note that $A^j\times\pi$ cannot enter a state in $Win_1(G_{\pi,j})$, since the $j$-queries failed. Thus, $A^j\times\pi$  stays in $Win_0(G_{\pi,j})$.
%While running on the primary trace, the goal automaton $A^j$ cannot enter a state $q$ $Win_1(G_{\pi,j})$, since the $j$-queries failed. 
A $j$-Deviant-Trace must separate from the primary trace at some time step $k\geq 0$, since deviant traces cannot be the primary trace, so at that point $A^j\times\pi$ is in some state $\langle q,s\rangle \in Win_0(G_{\pi,j})$. For $A^j$ to accept on a deviant trace means that agent 1 can force reaching, in the games $G_{\pi,j}$, from $\langle q,s\rangle$ to some $\langle q',s'\rangle$ for $q \in F^j$ and $s\in S$. But that is not possible, since it would mean that $\langle q,s\rangle \in Win_1(G_{\pi,j})$. It follows that $A^j$ cannot accept on a $j$-deviant trace.
%At this point, the agent $j$ must deviate to a state in $Win_0(G_{\pi,j})$ since they were previously in a state in $Win_0(G_{\pi,j})$ on the primary trace (otherwise a $j$-query would succeed), from which there is no winning strategy for agent $1$. We conclude that no $j$-Deviant-Trace satisfying $A^j$ can exist, and so the $j$-Deviant-Trace Condition is satisfied. 
\end{proof}

\paragraph{Complexity}
As noted before, each safety game and reachability query can be solved in PSPACE. Therefore, the entire algorithm has a PSPACE upper bound. The same holds for the bounded-channel case (for both goal automata and Moore machines), as the safety games and reachability queries would still be solved in PSPACE since $S$ would still be exponential in the size of the input.

\begin{theorem}
The verification problem with DFA goals can be solved in PSPACE.
\end{theorem}

\subsection{NFA and AFA Goals}
\subsubsection{NFA Goals}
The algorithm for NFA goals follows from the algorithm for DFA goals with some adaptation. Since we are dealing with nondeterministic automata now, we denote the transition function $\delta^i$ of the goal automaton $A^i$ as a set of triples with $\langle q, \alpha, q' \rangle$ belonging to $\delta^i$ if it possible to transition from state $q$ to  state $q'$ upon reading $\alpha \in \Sigma$.  As before, we start by considering the agents in $W$.

\paragraph{Checking Agents in $W$}
Given a goal automaton $A^i=(Q^i, q^i_0, \Sigma,\delta^i,  F^i)$ we use essentially the same construction of $A^i \times \pi = (Q^i \times S, \langle q_0^i, s_0 \rangle, \emptyset, \tau^i, F^i \times S)$, which is now a nondeterministic finite automaton. The transition function $\tau^i$ is modified slightly to accommodate nondeterminstic transitions. As before, the alphabet of this automaton is empty so transitions are defined between states; therefore  $\tau^i$ is represented a set of pairs. For $q\in Q^i$ and $s\in S$, We have $\langle \langle q, s \rangle, \langle q', s' \rangle \rangle \in \tau^i$ if $\langle q, \gamma(s), q' \rangle \in \delta^i$ and $s'=\rho(s,\gamma(s))$. Once again we test these automata for nonemptiness, noting that a word accepted by $A^i \times \pi$ corresponds to $A^i$ accepting a finite prefix of the primary trace of $\pi$. As before, we denote these nonemptiness queries as the i-queries. They can once again be tested for nonemptiness in NPSPACE=PSPACE, as they are once again equivalent to reachability testing in an exponentially large graph (caused by the exponential state space $S$).

As before, we proceed with the construction of safety games to analyze the set of states from which deviation is possible for an agent $j \not \in W$.

\paragraph{Safety Game for Deviating Agents}
We construct the safety  game $G_{\pi,j} = (Q^j\times S, Q^j \times S\times \Sigma, E_{\pi,j})$. agent 0 owns $Q^j\times S$, and agent 1 owns $Q^j\times S\times \Sigma$. The edge relation $E_{\pi,j}$ is defined as follows:
\begin{enumerate}
\item 
$(\langle q,s \rangle, \langle q,s,\alpha\rangle)) \in E_{\pi,j}$ for $q \in Q^j\setminus F^j$, $s\in S$, and $\alpha =\gamma(s)$.
\item 

$(\langle q,s,\alpha\rangle, \langle q',s'\rangle) \in E_{\pi,j}$ for $q,q'\in Q^j$ and  $s,s'\in S$, if $\langle q, \beta,  q' \rangle \in \delta^j$ 
and $s'=\rho(s,\beta)$, for some $\beta\in \Sigma$ such that $\alpha[-j]=\beta[-j]$.
\end{enumerate}

This is a slight modification from the previous construction that takes into account that there are now multiple transitions possible for a state $q$ and a letter $\beta \in \Sigma$ in $A^j$, so the fundamental structure of the game is unchanged. It still amounts to a reachability query, as agent 0 still has no choice in moves. As before, these safety games are exponential in the size of the input due to the presence of $S$, and therefore they can be solved in NPSPACE=PSPACE. We retain the notation that the set of winning states for agent 0 is given by $Win_0(G_{\pi,j})$ with $Win_1(G_{\pi,j})$ defined analogously.

\paragraph{Checking the Agents in $\Omega \setminus W$}
Once again, the same argument from before applies. We create the NFA $A^j \times \pi = ( Q^j \times S, \langle q_0^j, s_0 \rangle, \emptyset, \tau^j, Win_1(G_{\pi,j}) )$ from the goal NFA $A^j$, which differs from the previous construction of the NFA $A^i \times \pi$ in only the set of final states. As before, in the DFA case, we denote nonemptiness queries of $A^j \times \pi$ as the j-queries and they can once again be conducted in NPSPACE=PSPACE. It is once again integral to $\pi$ being a $W$-NE that the j-queries fail. We state an equivalent theorem to
Theorem~\ref{DFAmethod} for NFA inputs. 
\begin{theorem}
Given an iBG $G$ with NFA goal inputs, a strategy profile $\pi$ is a $W$-NE iff the $i$-reachability queries succeed and the $j$-reachability queries fail.
\end{theorem}

\begin{proof}
The proof of this theorem closely follows the proof of Theorem~\ref{DFAmethod}  and is therefore omitted.
\end{proof}

\paragraph{Complexity}
 Each safety game and reachability query was conducted in PSPACE. Therefore, the entire algorithm has a PSPACE upper bound. Once again, $S$ is exponential in the size of the input for even the bounded-channel case, so the result holds for the bounded-channel case as well.

\begin{theorem}
The verification problem with NFA goals can be solved in PSPACE.
\end{theorem}

We note that we can achieve the same upper bound by simply determinizing each goal automaton and then applying the DFA verification procedure. While both a DFA-based approach and the approach outlined in this section lie in PSPACE, the latter has better complexity in practice since it avoids a second exponential blowup. The approach in this section also prepares us to handle AFA goals. %in the next section.

\subsubsection{AFA Goals} 
A similar version of this problem in which a game $G$ with $LDL_f$ goal specifications was queried to see if some $W$-NE existed was presented in \cite{GPW17} and was proven PSPACE-complete. In this section, we show that this upper bound also holds when $W$ is specified and we are given AFA goals.

With AFA goals, we have a choice of converting to NFAs, incurring an exponential blowup, or DFAs, incurring a doubly exponential blowup. By converting to NFAs, we can avoid a second exponential blow up and show that this problem lies in PSPACE. Therefore given goal AFAs, we create equivalent NFAs $A^i$ from the input and proceed as before.

The safety games constructed for NFA goals had a state space of $\{Q^j\times S\} \cup \{Q^j \times S\times \Sigma \}$.  Since we converted from an AFA to NFA to obtain $Q^j$, $Q^j$ is now exponential in the size of the input. Since $S$ remains exponential and $\Sigma$ was a part of the input, this game remains exponential in the size of the input. Therefore, these safety games can still be solved in PSPACE.

With respect to the automata constructed for the reachability queries, each vertex space $Q^i \times S$ still remains exponential in the size of the input even when $Q^i$ is exponential in the size of the input. Therefore, these reachability queries can also be solved in PSPACE.

\begin{theorem}
The verification problem with AFA goals can solved in PSPACE.
\end{theorem}

Note that by employing an approach in which we check each agent individually, we have also avoided a situation in which we must take the cross product of exponentially large automata. If we are given $k$ AFAs and wish to convert them in $k$ NFAs, this represents an exponential blowup. At this point, we could take the cross product of all NFAs and still remain exponential in the size of the input. Note, however, that this would involve a  quadratic blowup in the exponent - while $2^{n}$ and $2^{kn}$ are both exponential in $n$, there is an exponential (in $k$) gap between $2^n$ and $2^{kn}$.

%\sr{Bounded Channel Automata}

\subsection{Lower Bound}
We now provide a PSPACE-hardness result for the verification problem with DFA goals, which also serves as a lower bound for the verification problems with NFA and AFA goals.  
We use the succinct representations of bounded-channel automata to show that the verification problem for DFA goals is PSPACE-hard through a polynomial time reduction from the following canonical PSPACE-complete problem : given a deterministic Turing machine $M$ and a natural number $n$ in unary, does $M$ accept the empty tape using at most $n$ space~\cite{sipser2006}?  We further assume that $M$ has a unique accepting configuration in which the tape consists solely of a special unused character $*$ with the head on the rightmost of the $n$ cells. This standard assumption does not influence the complexity of the problem. 
%as modifying any Turing machine to one that accepts the same language with such an accepting configuration can be done in constant time.

The Turing Machine $M$ has a state set denoted by $R$ and an alphabet denoted by $\Delta$. Our reduction relies on the notion of an instantaneous description (ID) of a Turing Machine, which is a string that represents the content of the tape at a discrete time step in the run time of $M$. Such an ID includes
\begin{itemize}
\item The complete contents of the tape from left to right.
\item The position and state of the head of $M$. As a matter of notation, if the head is on cell~$i$ then the character corresponding to the content of cell~$i$ is a pair consisting of the the element of $\Delta$ on the tape \emph{and} an element of $R$ representing the state of the machine.
\end{itemize}
As an example, an ID could be of the form $121\langle0,q\rangle31$. In this case, the content of the tape is $121031$, while the pair $\langle0,q\rangle$ denotes that the machine is currently reading the cell with symbol $0$ while in state $q$. Since the machine is deterministic, a sequence of IDs corresponding to the computation run of $M$ on the empty tape is uniquely given by the initial state and position of the head of $M$, which we will call $ID_0$. The machine then accepts if there is a sequence of IDs $ID_0 \ldots ID_m$ such that $ID_m$ is the unique accepting configuration of $M$ and $ID_{i+1}$ follows from $ID_i$ according to the transition function  of $M$.
Our reduction strategy to the verification problem is to use a set of bounded-channel Moore Machines to simulate the transitions from one $ID_i$ to $ID_{i+1}$, and a set bounded channel DFA goals of the agents to verify that the sequence is correct - that it starts at $ID_0$ and eventually reaches $ID_m$.

We now sketch a construction of a game $G_{M,n}$ and a strategy profile $\pi_{M,n} = \langle \pi_0 \ldots \pi_{n-1} \rangle $ such that the Turing machine $M$ accepts the empty tape in at most $n$ steps iff $\pi$ is an $\Omega$-NE in $G$, i.e. a strategy profile that satisfies every agent's goal on its primary trace. The number of agents in this game is given by $n$, the same as the length of the unary input to the Turing Machine $M$.

We first consider the $\Sigma_i$ assigned to each agent in $G$. This alphabet is $R \cup \{ \Delta \times R \}$ for every agent. Intuitively, each $\Sigma_i$ represents a single character of the ID, with $\Sigma_i$ specifically corresponding to the $i$-th cell of the ID. Therefore taking all $n$ alphabets together as $\Sigma$ corresponds to an entire ID.  

The strategy $\pi_i$ of an agent $i$ outputs the next configuration of cell~$i$ based on the previous configurations of the cell to the right, the cell to the left, and the cell itself. Thus, each strategy only needs to read at most three symbols, from $\Sigma_{i-1},\Sigma_i$, and $\Sigma_{i+1}$--since transitions in a Turing machine are determined locally--and output a symbol in $\Sigma_i$ according to the transition function of $M$. (Agents $0$ and $n-1$ need only consider two of the $\Sigma_i$ since there are no cells to the left and right of these agents, respectively). If the computation moves out of bounds, then the Moore machine for the agent that moved it out of bounds (either $0$ or $n-1$) immediately moves to a sink state that continuously outputs a special character that does not appear in the unique accepting configuration. Since each strategy only considers at most three $\Sigma_i$'s  as input and outputs a single $\Sigma_i$, these strategies can be represented by bounded-channel Moore machines. The state space of each machine has a size upper bound of  $|R \cup \{ \Delta \times R \}|^3$, meaning that each machine is polynomial in the size of the input.

We now consider the goals for the agents. The purpose of the goals is to check that the initial ID represent the empty tape, and the final ID is an accepting one. The goal for each agent except Agents $0$ and $n-1$ is to see the symbol $*$ some time after reading the empty symbol $\bot$ as the first symbol corresponding to the empty tape input. The goal for Agent 0 is to eventually read the symbol $*$, after reading the first symbol $\langle \bot,q_0\rangle$, where $q_0\in R$ is the initial state of $M$. The goal for Agent $n-1$ is to read the pair $\langle *, q_F \rangle$ some time after seeing an empty cell as the first character, where $q_F \in R$ is the unique state corresponding to the accepting configuration $ID_m$ since the head is moved all the way to the right in $ID_m$. The DFA representations of these goals are very simple, as they solely consist of eventually reading a single character after verifying an initial character. Therefore, all goal DFAs are bounded-channel automata and are therefore polynomial in the size of the input. Overall, we have a polynomial number of agents, each with a polynomial-sized Moore machine and a polynomial-sized goal DFA. Therefore, the game $G$ can be constructed in polynomial time. We now prove the correctness of the reduction.

\begin{theorem}
The strategy profile $\pi_{M,n} = \langle \pi_0 \ldots \pi_{n-1} \rangle$ is an $\Omega$-NE in $G_{M,n}$ iff $M$ accepts the empty tape using at most space $n$. 
\end{theorem}
\begin{proof}
$(\rightarrow)$  Assume that $\pi_{M,n}$ is an $\Omega$-NE. By construction, the primary trace of $\pi_{M,n}$ simulates the sequence of IDs of $M$ running on the empty tape with a built in check to ensure that the computation uses no more than $n$ cells. Therefore, for the DFA goals to accept on this trace it means that in the final ID all cells except the last are filled with the special $*$ character and the last has the pair $\langle *, q_F \rangle$ and that each cell started empty. This means that there is a valid sequence of IDs generated by $M$ upon reading the empty tape, meaning that $M$ accepted the empty tape while staying in bounds.
%-  if it had then it would have no way of transitioning to $q_F$.

$(\leftarrow)$Assume that $M$ accepts the empty tape using no more than $n$ space. Then, it generates a  unique valid sequence of IDs that eventually end at the unique accepting configuration. By construction, the primary trace of $\pi$ consists of this same sequence of IDs, and $ID_m$ must consist of the unique configuration consisting of $*$ on every cell but the rightmost with the pair $\langle *, q_F \rangle$. Therefore, all DFA goals will accept on the primary trace of $\pi_{M,n}$, so $\pi_{M,n}$ is an $\Omega$-NE.
%If any DFA goal has not accepted yet, upon seeing $ID_m$ they must as this is how the goals were defined. Since all DFAs accept on the primary trace of $\pi$, $\pi$ is an $\Omega$-NE by definition.
\end{proof}

We are able to construct the game $G_{M,n}$ and the profile $\pi_{M,n}$ in polynomial time due to the succinct representation of bounded-channel automata. Therefore, we have exhibited a polynomial time reduction from a known PSPACE-complete problem for the verification problem with DFA goal inputs. Combining this with our PSPACE upper bound, we get
\begin{theorem} The verification problems with DFA goals is PSPACE-complete.
\end{theorem}

Since DFAs are a special case of both NFAs and AFAs, we get lower bounds for both corresponding verification problems as well.
\begin{corollary} 
The verification problems with NFA or AFA goals are PSPACE-complete.
\end{corollary}

\section{Concluding Remarks}

In this work we provided complexity results for both the realizability and verification problems in the finite-horizon multiagent setting with different types of goal specifications, significantly extending previous works \cite{GPW17,RV21}. One of the key points of interest from this analysis is the complexity gap observed between the complexities of the realizability and verification problems. While realizability with DFA goals was proven to be PSPACE-complete in \cite{RV21}, here we have shown that realizability with NFA goals is EXPTIME-complete and realizability with AFA goals is 2EXPTIME-complete. Therefore, with respect to the realizability problem, we have shown that the succincness of goal specification greatly influences the complexity of the realizability problem. With respect to the verification problem, however, this distinction does not exist, as the verification problems with DFA,NFA, and AFA goals are all PSPACE-complete. Thus, the complexity for DFA goals is the same for both the realizability and the verification problems, but as the automata get more succinct the realizability problem grows in complexity while the verification problem remains PSPACE-complete. This complexity picture is similar to what is known in temporal reasoning in two-agent systems (system and environment), where the complexity of realizability rises from PTIME for DFA goals to 2EXPTIME for $LTL_f$ goals, while verification, i.e., model checking, is PSPACE-complete for different types of goals, with the system \emph{state-explosion problem} being the primary source of PSPACE-hardness, cf.~\cite{Var96}
%Interestingly enough, it seems that this phenomenon is largely driven by the game constructed for deviating agents, even though PSPACE-hardness for the realizability problem with DFA goals was shown through a game with no deviating agents (i.e. $W=\Omega$) \cite{RV21}.

Finally, by reasoning about the Primary-Trace Condition and the $j$-Deviant-Trace Conditions separately, as in \cite{RV21}, we were able to get algorithms that are easy to understand and optimal. This method of separation was made specifically to reason about Nash equilibria in qualitative games, by leveraging properties of both the Nash equilibria as a solution concept and qualitative goals themselves. By analyzing the $j$-Deviant-Trace Condition separately through the use of safety games we are able to get much better complexity bounds than if we dealt with the entire Nash equilibria at once.  We believe that this principle of separation provides a powerful framework to reason about other qualitative multi-agent systems and perhaps even other solution concepts.

%\myv{Please be clear, not generous!}

\section*{Acknowledgements}
Work supported in part by NSF grants IIS-1527668, CCF-1704883,
IIS-1830549, CNS-2016656, DoD MURI grant N00014-20-1-2787,
and an award from the Maryland Procurement Office.
\bibliographystyle{kr} 
\bibliography{references}

\end{document}